\documentclass{iopconfser}

\usepackage{graphicx,cite}

\begin{document}

\title{Symmetries and singular behaviors with Bohmian trajectories}

\author{\'Angel S. Sanz}

\affil{Department of Optics, Faculty of Physical Sciences, Universidad Complutense de Madrid\\ Pza.\ Ciencias 1, 28040 Madrid, Spain}

\email{a.s.sanz@fis.ucm.es}

\begin{abstract}
Quantum mechanics is able to predict challenging behaviors even in the simplest physical scenarios.
These behaviors are possible because of the important dynamical role that phase plays in the evolution of quantum systems, and are very similar, on the other hand, to effects observable in analogous optical systems.
This work focuses on how Bohmian mechanics proves to be a rather convenient theoretical framework to analyze phase-based phenomena, since the phase constitutes the central element in this hydrodynamical formulation of quantum mechanics.
More specifically, it allows us to understand how spatial phase variations give rise to velocity fields that eventually rule the dynamical behavior of quantum systems, and that, when integrated in time locally (i.e., taking into account specific positions), they provide us with a neat local (point by point) description of the system evolution in the configuration space.
Indeed, it will also be seen that this idea transcends the quantum realm and can be profitably used to describe the behavior of optical analogs with rather singular behaviors.
With this purpose, two interesting phenomena that take place in free space are considered, namely, the self-acceleration and shape-invariance of Airy beams, and spontaneous self-focusing.
\end{abstract}

%%%%%%%%%%%%%%%%%%%%%%%%%%%%%%%%%%%%%%%%%%%%%%%%%%%%%%%%%%%%%%%%%%%%%%%%
%%%%%%%%%%%%%%%%%%%%%%%%%%%%%%%%%%%%%%%%%%%%%%%%%%%%%%%%%%%%%%%%%%%%%%%%

\section{Introduction}
\label{sec1}

Since its formulation by David Bohm in 1952 \cite{bohm:PR:1952-1,bohm:PR:1952-2}, Bohmian mechanics has been widely understood as a causal interpretation or model for quantum phenomena, alternative to the current quantum theory.
With its seminal work, Bohm opened up a new way to conceive quantum physics, aligned with other alternative approaches aimed at favoring a {\it quantum theory without observers} \cite{goldstein:phystoday:1,goldstein:phystoday:2}.
Yet, Bohmian mechanics can also be simply considered, from a strictly formal point of view, a direct consequence of reformulating standard quantum mechanics in hydrodynamic terms, as it was done by Erwin Madelung in 1926 \cite{madelung:ZPhys:1926}, which also leads to interesting implications and applications \cite{holland-bk,wyatt-bk,sanz-bk-1,sanz-bk-2,benseny:EPJD:2014}.
In this case, the concept of probability current or quantum flux \cite{bohm-bk:QTh,schiff-bk} acquires a major role, since it constitutes a direct link that connects the usual evolution of the probability density in the configuration space, described by the Schr\"odinger equation, with the so-called Bohmian momentum or guidance condition.
In this way, Bohm's momentum does not need to be considered an external concept to the usual quantum theory; we find that this concept is indeed contained in a natural fashion within this theory \cite{sanz:FrontPhys:2019}.
Indeed, physically, it is a measure of the local spatial variations undergone by the phase, while the so-called Bohmian trajectories constitute, in turn, a local measure of the flux along precise streamlines, which have nothing to do with real trajectories followed by a quantum system \cite{sanz:AJP:2012,sanz:JPhysConfSer:2012,sanz:JPhysConfSer:2014}.

~

Following the above discussion, we can thus understand the so-called Bohmian trajectories either as hidden variables, providing us with an alternative perspective of quantum systems \cite{bohm-bk:1980,bohm-hiley-bk,cushing-bk:1994,cushing-bk:1996,holland-bk,duerr-bk:2013,bricmont-bk,norsen-bk}, or, alternatively, as a methodological tool, worth applying to a wide variety of physical problems to get new insights \cite{wyatt-bk,sanz-bk-1,sanz-bk-2,benseny:PanStanford:2012,benseny:EPJD:2014}, which may have either a quantum mechanical nature, or necessarily not.
Here, we are going to focus on the second point of view, trying to settle in simple terms what a Bohmian trajectory is \cite{sanz:JPhysConfSer:2012,sanz:JPhysConfSer:2014} and what it can be good for \cite{sanz:FrontPhys:2019}, leaving aside expendable ontological connotations.
This will also serve us to show how this quantum trajectory-based methodology can be extrapolated to other non-quantum contexts, which keep a certain resemblance at a formal level.

~

To illustrate such claims, here two challenging wave phenomena in free space, directly associated with the phase, are analyzed and discussed by means of Bohmian mechanics: (1) solutions that contain an infinite amount of energy, with particular emphasis on the phenomenology of Airy wave packets due to their self-acceleration and shape-invariance properties \cite{sanz:JOSAA:2022}, and (2) self-focusing solutions \cite{sanz:PLA:2024}, i.e., solutions that converge to a singularity with time without the action of any external agent or nonlinear contribution \cite{peres-bk,aiello:OptLett:2016,aiello:OSACont:2020,porras:PRA:2021}.
In both cases, it will be shown how, despite the peculiar nature of these solutions, they admit well-posed trajectory-based descriptions, which help us to understand in an intuitive manner the main features (the challenging ones) of their respective dynamical evolutions and, from them, the major role played in those dynamics by the local variations of the quantum phase. An interesting point here is thus the one concerning the symmetries displayed by the underlying velocity fields (or, equivalently, the associated phase fields), and hence the physical implications arising when such symmetries are broken.

~

The work is organized as follows.
In Sec.~\ref{sec2} a brief overview of the main theoretical aspects related to Bohmian mechanics is summarized, as well as its extension to standard paraxial wave optics, which relies on a Schr\"odinger-type equation, namely, the paraxial Helmholtz equation.
Section~\ref{sec3} presents a discussion of the Airy beam propagation in terms of Bohmian trajectories, while spontaneous self-focusing is considered in Sec.~\ref{sec4}.
Finally, a series of concluding remarks are highlighted in Sec.~\ref{sec5}.

%%%%%%%%%%%%%%%%%%%%%%%%%%%%%%%%%%%%%%%%%%%%%%%%%%%%%%%%%%%%%%%%%%%%%%%%
%%%%%%%%%%%%%%%%%%%%%%%%%%%%%%%%%%%%%%%%%%%%%%%%%%%%%%%%%%%%%%%%%%%%%%%%

\section{Bohmian trajectories as a phase-mediated-dynamics probing tool}
\label{sec2}

%%%%%%%%%%%%%%%%%%%%%%%%%%%%%%%%%%%%%%%%%%%%%%%%%%%%%%%%%%%%%%%%%%%%%%%%

\subsection{Velocities and trajectories in standard quantum mechanics}
\label{sec21}

Consider a spinless, nonzero mass quantum system described in the configuration space, at a time $t=0$, by the wave function $\psi({\bf r},0)$.
If the system is in free space, the time evolution of the system wave function at any other later time is ruled by the simple time-dependent Schr\"odinger equation
\begin{equation}
	i\hbar \frac{\partial \psi({\bf r},t)}{\partial t} = - \frac{\hbar^2}{2m} \nabla^2 \psi ({\bf r},t) .
	\label{eq1}
\end{equation}
If this equation is multiplied by the conjugate field variable, $\psi^*({\bf r},t)$, and then we subtract (to the resulting equation) the conjugate equation, we readily obtain the continuity equation
\begin{equation}
	\frac{\partial \rho({\bf r},t)}{\partial t} = - \nabla \cdot {\bf j}({\bf r},t) ,
	\label{eq2}
\end{equation}
where $\rho({\bf r},t) = |\psi({\bf r},t)|^2$ is the usual probability density and
\begin{equation}
	{\bf j}({\bf r},t) = \frac{\hbar}{2mi} \Big[ \psi^*({\bf r},t) \nabla \psi({\bf r},t) - \psi({\bf r},t) \nabla \psi^*({\bf r},t) \Big]
	\label{eq3}
\end{equation}
is the probability current or quantum flux \cite{bohm-bk:QTh,schiff-bk}.
Since Eq.~(\ref{eq2}) is a transport equation, the quantum flux (\ref{eq3}) can be rewritten as ${\bf j}({\bf r},t) = {\bf v}({\bf r},t) \rho({\bf r},t)$, where the transport velocity field is
\begin{equation}
	{\bf v}({\bf r},t) = \frac{{\bf j}({\bf r},t)}{\rho({\bf r},t)} = \frac{\hbar}{m} {\rm Im} \left[ \frac{\nabla \psi({\bf r},t)}{\psi({\bf r},t)} \right] = \frac{1}{m} {\rm Re} \left[ \frac{\hat{\bf p} \psi({\bf r},t)}{\psi({\bf r},t)} \right]
	\label{eq4}
\end{equation}
and $\hat{\bf p} = -i\hbar\nabla$ is the usual momentum operator.
Accordingly, the quantity ${\bf p}({\bf r},t) = m{\bf v} ({\bf r},t)$ can be understood as a measure of the local value of the momentum, complementary of the corresponding expectation value, which provides us with a sort of average of ${\bf p}$ over the whole configuration space.

~

So far, all concepts introduced above are included within standard quantum mechanics and, therefore, there is no need to consider extra postulates.
Let us now briefly overview the conventional Bohmian approach, which merges the standard quantum theory with the concept of hidden variable, i.e., assigning specific positions in configuration space to quantum systems \cite{bohm:PR:1952-1,bohm:PR:1952-2,holland-bk,duerr-bk:2009}.
The evolution in time of these positions give rise to well-defined trajectories, known as Bohmian trajectories, which do not violate any fundamental law of quantum mechanics, such as the uncertainty relations or the complementarity principle.
To reach such a trajectory-based formulation, Bohm considered \cite{bohm:PR:1952-1,holland-bk} the nonlinear transformation relation $\{\psi, \psi^*\} \to \{\rho, S\}$, from two complex field variables to two real field variables, such that
\begin{equation}
	\psi({\bf r},t) = \sqrt{\rho({\bf r},t)} e^{i S({\bf r},t)/\hbar} ,
	\label{eq5}
\end{equation}
and the same for $\psi^*$.
After making the corresponding substitution into Eq.~(\ref{eq1}), and its conjugate, and then separating the real and imaginary parts of the resulting equation, one obtains the equations of motion for both $\rho({\bf r},t)$ and $S({\bf r},t)$, namely,
\begin{eqnarray}
	\frac{\partial \rho({\bf r},t)}{\partial t} & = & - \nabla \cdot \left[ \rho({\bf r},t) \frac{\nabla S({\bf r},t)}{m} \right] ,
	\label{eq6} \\
	\frac{\partial S({\bf r},t)}{\partial t} & = & \frac{\left[ \nabla S({\bf r},t) \right]^2}{2m} + Q({\bf r},t) .
	\label{eq7}
\end{eqnarray}
Equation~(\ref{eq6}) corresponds to the continuity equation (\ref{eq2}), although the velocity field appears recast in terms of the new phase variable, i.e., ${\bf v} = \nabla S/m$, while Eq.~(\ref{eq7}) is the so-called quantum Hamilton-Jacobi equation, which generalizes the classical version \cite{holland-bk,goldstein-bk} by including the extra contribution $Q({\bf r},t)$, known as {\it Bohm's potential} or {\it quantum potential},
\begin{equation}
	Q({\bf r},t) = - \frac{\hbar^2}{2m} \left\{ \frac{\nabla^2 \rho({\bf r},t)}{\rho({\bf r},t)} - \frac{1}{2} \left[ \frac{\nabla \rho({\bf r},t)}{\rho({\bf r},t)} \right]^2 \right\} .
	\label{eq8}
\end{equation}
Although $Q$ is commonly referred to as a potential, it should be noticed that this contribution has nothing to do with a potential energy function, but with the original kinetic operator $\hat{\bf p}/2m = -\hbar^2 \nabla^2/2m$.
Because we are considering free space propagation, in Eq.~(\ref{eq7}) we explicitly see, without the contamination of an external potential function, that the kinetic operator gives rise to two kinetic contributions, one directly related with the phase of the wave function, and another one related to the curvature of its amplitude (or, equivalently, of its associated probability density).

~

Given the similarity between Eq.~(\ref{eq7}) and the classical Hamilton-Jacobi equation, Bohm postulated \cite{bohm:PR:1952-1,holland-bk} the existence of trajectories supplementing the description of the quantum system already provided by the wave function.
These trajectories, which would not be observable (thus being hidden variables), are obtained after integrating in time the equation of motion
\begin{equation}
	\dot{\bf r}({\bf r},t) = \frac{{\bf p}({\bf r},t)}{m} = \frac{\nabla S({\bf r},t)}{m} ,
	\label{eq9}
\end{equation}
where ${\bf p}$ is the so-called {\it Bohm's momentum}, which corresponds to the quantity ${\bf p} = m {\bf v}$ introduced above.
From a formal point of view, this equation of motion, known as the {\it guidance condition} or {\it equation}, tells us how a local velocity field, intrinsically associated with the spatial variations of the phase, rules the distribution of the probability density in configuration space.
Maybe this is a rather challenging consequence in quantum mechanics.
However, note that in optics the engineering of structured light beams \cite{forbes:NatPhotonics:2021}, which precisely follow similar equations (see below), is actually possible by making light to pass through devices that produce such spatial phase variations (i.e., spatial light modulators).
Even more, with the possibility to experimentally obtain {\it weak measurements} \cite{aharonov:PRL:1988,sudarshan:PRD:1989}, it has been possible to measure this transverse field and, from it, to infer the path followed by the corresponding trajectories \cite{wiseman:NewJPhys:2007,kocsis:Science:2011,steinberg:SciAdv:2016} (which has nothing to do with the hypothetical pathways that could be expected to be followed by individual photons).

~

From a conventional point of view, we thus have a combination of standard tools in a hydrodynamic language (\`a la Madelung \cite{madelung:ZPhys:1926}) and a postulated momentum (\`a la Bohm \cite{bohm:PR:1952-1}), which does not disprove the above pragmatic consideration of Bohmian mechanics as a valid methodology or quantum representation, beyond its commonly accepted conception of being a hidden-variable model.
This leads us to a third aspect, which emphasizes the usefulness of this approach as a convenient computational tool.
Specifically, when one deals with the issue of propagating the wave function following the Schr\"odinger equation, there are different strategies, the so-called spectral methods being a widely used methodology \cite{feit-fleck:JCompPhys:1982,feit-fleck:JCP:1983,kosloff:JCompPhys:1983}.
The basic idea of these methods is to recast the wave function as a linear superposition of eigenfunctions of the Laplacian, so that the nonlocal action of the kinetic term can be evaluated in a more accurate manner than with discretizing methods (where the accuracy of the method is associated with the number of neighbors included in the approximation of the discrete Laplacian).
From a Bohmian point of view, this is a rather convenient strategy to obtain analytical expressions for the equation of motion in terms of the spectral or pseudo-spectral components of the wave function \cite{sanz-bk-2}.
Indeed, if the method is based on the Fourier transform, these components can directly be the  components of the discrete Fourier transform, which simplifies remarkably the numerical calculation of the trajectories and, at the same time, also increases their numerical accuracy (something very important in problems characterized by the presence and sudden emergence of nodes).

~

Consider, for simplicity, that the wave function can be decomposed as a sum of plane waves, i.e.,
\begin{equation}
	\psi({\bf r},t) = \sum_{j=-n}^n c_j e^{{\bf k}_j\cdot {\bf r} - E_j t/\hbar} ,
	\label{eq10}
\end{equation}
where $c_j = |c_j| e^{\phi_j}$, ${\bf k}_j = {\bf p}_j/\hbar$, and $E_j = {\bf p}_j^2/2m$.
Note that the ${\bf k}_j$ might correspond, for instance, to the discrete set of momenta that arise as soon as the configuration space is discretized in a mesh of equidistant knots, when we apply the fast Fourier method \cite{press-bk-2}.
If Eq.~(\ref{eq10}) is substituted into Eqs.~(\ref{eq4}) or, equivalently, (\ref{eq9}), we obtain the equation of motion
\begin{equation}
	\dot{\bf r}({\bf r},t) = \frac{\hbar}{m} \frac{\sum_{j,l=-n}^n |c_j| |c_l| {\bf k}_j \cos (\Delta {\bf k}_{jl} \cdot {\bf r} - \omega_{jl} + \delta_{jl})}{\sum_{j,l=-n}^n |c_j| |c_l| \cos (\Delta {\bf k}_{jl} \cdot {\bf r} - \omega_{jl} + \delta_{jl})} ,
	\label{eq11}
\end{equation}
where $\Delta {\bf k}_{jl} = {\bf k}_j - {\bf k}_l$, $\omega_{jl} = (E_j - E_l)/\hbar$, and $\delta_{jl} = \phi_j - \phi_l$.
If the initial wave function is now recast in terms of its spectral decomposition in momenta as
\begin{equation}
	\psi({\bf r},0) = \sum_{j=-n}^n c_j e^{{\bf k}_j\cdot {\bf r}} = \sum_{j=-n}^n \tilde{\psi}({\bf k}_j,0) e^{{\bf k}_j\cdot {\bf r}} ,
	\label{eq12}
\end{equation}
then Eq.~(\ref{eq11}) reads as
\begin{equation}
	\dot{\bf r}({\bf r},t) = \dot{\bf r} \left[ \tilde{\psi}({\bf k},0), {\bf r},t \right] = \frac{\hbar}{m} \frac{\sum_{j,l=-n}^n |\tilde{\psi}({\bf k}_j,0)| |\tilde{\psi}({\bf k}_l,0)| {\bf k}_j \cos (\Delta {\bf k}_{jl} \cdot {\bf r} - \omega_{jl} + \delta_{jl})}{\sum_{j,l=-n}^n |\tilde{\psi}({\bf k}_j,0)| |\tilde{\psi}({\bf k}_l,0)| \cos (\Delta {\bf k}_{jl} \cdot {\bf r} - \omega_{jl} + \delta_{jl})} .
	\label{eq13}
\end{equation}

~

Summarizing, we can see that, depending on whether one wishes to explore global or local properties of the quantum system, there are two supplementing elements worth keeping in mind.
For global properties, the wave function and the properties obtained from it in the usual manner play an important role.
On the other hand, if we wish to understand what happens dynamically at a local level, i.e., why the time evolution of the system is of one or another form, which involves phase information, although the latter is still contained within the wave function, the suitable probing tool here will be either considering the velocity field or the trajectories obtained after integrating it.
This will be illustrated by means of the two systems with interesting symmetries, which are being considered below in Secs.~\ref{sec2} and~\ref{sec3}.

%%%%%%%%%%%%%%%%%%%%%%%%%%%%%%%%%%%%%%%%%%%%%%%%%%%%%%%%%%%%%%%%%%%%%%%%

\subsection{From trajectories in standard quantum mechanics to rays in standard scalar optics}
\label{sec22}

Although the Bohmian methodology is commonly regard to quantum mechanics, it should be noted that it is quite general, as it will also apply to any other physical system described by analogous equations.
It is always interesting to note the similarity among different physical problems, particularly because precisely, quoting Feynman, ``the equations for many different physical situations have exactly the same appearance'', which means that ``having studied one subject, we immediately have a great deal of direct and precise knowledge about the solutions of the equations of another'' \cite{feynman:FLP2:1965}.
In this sense, since we are interested in applying the Bohmian ideas to understand the shaping of light, let us consider the amplitude of an electromagnetic scalar field, $U({\bf r},t)$.
As it is well known, its propagation in time is described by the wave equation \cite{bornwolf-bk},
\begin{equation}
	\nabla^2 U({\bf r},t) - \frac{1}{v^2} \frac{\partial^2 U({\bf r},t)}{\partial t} = 0 ,
	\label{eq14}
\end{equation}
where $v = c/n$ is the field propagation speed in any direction, with $n$ being the refractive index of the medium that the field propagates through.
If the field is monochromatic, i.e., $U({\bf r},t) = \Psi({\bf r},t) e^{-i\omega t}$, the above equation becomes the time-independent Helmholtz equation,
\begin{equation}
	\nabla^2 \Psi({\bf r},t) + k^2 \Psi({\bf r},t) = 0 ,
	\label{eq15}
\end{equation}
where $v = c/n = \omega/k$.

~

Let us now assume that the energy propagates mostly along a given direction (say, $z$), while the effects of interest lie on perpendicular planes.
This means that we can assume paraxial conditions, i.e., all important effects take place close to the $z$-axis as the field mainly moves forward parallel to it.
Accordingly, we can write the time-independent field amplitude as $\Psi({\bf r}_\perp,z) \approx \psi({\bf r}_\perp,z) e^{ikz}$, i.e., as a plane wave accounting for the propagation along the $z$-axis, with a modulating amplitude along the transverse direction.
With this assumption, and after neglecting slow second order derivatives, Eq.~(\ref{eq15}) becomes the paraxial Helmholtz equation,
\begin{equation}
	i\frac{\partial \psi({\bf r},t)}{\partial z} = - \frac{1}{2z} \nabla_\perp^2 \psi({\bf r}_\perp,z) ,
	\label{eq16}
\end{equation}
which is formally isomorphic to Schr\"odinger's equation in free space.
Equation~(\ref{eq16}) physically tells us that, effectively, the transverse distribution of the light field can be understood as a propagation in terms of the (propagation) coordinate $z$.
Note that if we consider the variable transformation
\begin{equation}
	z = \frac{\hbar k}{m} t = \frac{\hbar n}{m\lambda_0} t ,
	\label{eq17}
\end{equation}
and $\nabla^2_\perp$ is replaced by $\nabla^2$, then we immediately recover Schr\"odinger's equation for propagation in free space.
Accordingly, considering the same transformation, from Eq.~(\ref{eq9}), we find that an analogous ray equation can also be obtained in optics, since
\begin{equation}
 \frac{m}{\hbar}\frac{d{\bf r}}{dt} \rightarrow k \frac{d{\bf r}_\perp}{dz} .
	\label{eq18}
\end{equation}
The above relation basically says that the momentum and the wave vector are linearly related one another, since, making both sides to be equal, we obtain ${\bf p}/\hbar = {\bf k}$, which is precisely {\it de Broglie's quantization condition}, which dynamically associates the particle aspect (${\bf p}$) with a wave aspect (${\bf k}$ or, equivalently, $1/\lambda$) --- also valid for photons if Einstein's equations for the energy of a zero-mass particle and for the photon energy are made equal.

~

Making the corresponding substitutions, we obtain an effective transverse velocity field or momentum,
\begin{equation}
	\frac{d{\bf r}_\perp}{dz} \equiv {\bf v}_\perp ({\bf r}_\perp,z) = \frac{1}{k} {\rm Im} \left[ \frac{\nabla \psi({\bf r}_\perp,z)}{\psi({\bf r},z)} \right] ,
	\label{eq19}
\end{equation}
where ${\bf v}_\perp$ is not a velocity in strict sense, but a field that behaves similarly in an ``effective'' manner.
This effective field is analogous to the transverse momentum introduced in \cite{sanz:AOP:2015} to describe the properties of an atomic Mach-Zehnder interferometer.
The integration of this quantity in $z$ provides us with trajectories that describe the spatial distribution of the electromagnetic energy or, equivalently, how the latter redistributes along the transverse direction as $z$ increases.
It is worth noting that the outcomes experimentally reported in \cite{kocsis:Science:2011} correspond precisely to the transverse field (\ref{eq19}), while the inferred trajectories are in compliance with what one would obtain by integrating this effective equation of motion.
On the other hand, it is also worth emphasizing that Eq.~(\ref{eq19}) for paraxial conditions is an approximation to the general expression that arises from the Poynting vector \cite{sanz:ApplSci:2020}.

~

We thus find that, regardless of whether we have a quantum system propagating in free space or an optical light beam propagating in a homogeneous and isotropic medium, we can use in their description the same equations to account for both the global and local properties, namely,
\begin{eqnarray}
	i \frac{\partial \psi({\bf u},\xi)}{\partial \xi} & = & - \frac{1}{2} \nabla_{\bf u}^2 \psi ({\bf u}, \xi) ,
	\label{eq20} \\
	{\bf v}_{\bf u}({\bf u},\xi) & = & {\rm Im} \left[ \frac{\nabla_{\bf u} \psi({\bf u}, \xi)}{\psi({\bf u}, \xi)} \right] ,
	\label{eq21}
\end{eqnarray}
where ${\bf u}$ refers to the spatial coordinates (transverse, in the case of the paraxial optical system), $\xi$ is the propagation variable, and $\nabla^2_{\bf u}$ and $\nabla_{\bf u}$ are, respectively, the Laplacian and the gradient computed with respect to the spatial coordinates ${\bf u}$, and ${\bf v}_{\bf u}$ denotes the effective velocity field in the space defined by the coordinates ${\bf u}$.
After integrating ${\bf v}_{\bf u}$ along the propagation variable $\xi$, for a set of initial conditions ${\bf u}_0 (\xi = 0)$, one obtains ensembles of trajectories parameterized in terms of the $\xi$, which provide us with an alternative description of the system propagation.

%%%%%%%%%%%%%%%%%%%%%%%%%%%%%%%%%%%%%%%%%%%%%%%%%%%%%%%%%%%%%%%%%%%%%%%%
%%%%%%%%%%%%%%%%%%%%%%%%%%%%%%%%%%%%%%%%%%%%%%%%%%%%%%%%%%%%%%%%%%%%%%%%

\section{Self-accelerating shape-invariant propagation}
\label{sec3}

A free-propagating Gaussian wave packet preserves its Gaussian amplitude in configuration space at any time, with its width undergoing a hyperbolic increase \cite{sanz-bk-2,sanz:AJP:2012}.
Accordingly, for $t \gg 2m\sigma_0^2/\hbar$ (with $\sigma_0$ being its width at $t = 0$), the amplitude essentially increases linearly with time, thus remaining scale invariant in spite of its progressive dispersion.
In 1979, Berry and Balazs \cite{berry:AJP:1979} showed that it is also possible to find non-dispersive wave packet solutions to Schr\"odinger's equation in configuration space.
The only requirement to observe this invariance property during their propagation is that the initial amplitude of the wave packet has to be described by an Airy function.
The amplitude of these so-called Airy wave packets thus presents two important symmetries: ($i$) shape invariance during propagation, unlike Gaussian wave packets, and ($ii$) self-acceleration without the presence of any external potential.
Almost thirty years later, Siviloglou {\it et al.}~\cite{christodoulides:PRL:2007} produced the first experimental evidence of this type of dispersionless wave packets with modulated light beams, making use of the isomorphism discussed in Sec.~\ref{sec22}.

~

To understand the properties of Airy beams, let us consider a one-dimensional beam with its initial wave function being described by an Airy function, i.e.,
\begin{equation}
	\psi(x,0) = Ai(x) ,
	\label{eq22}
\end{equation}
with
\begin{equation}
	Ai(s) = \frac{1}{\pi} \int_0^\infty \cos (u^3/3 + su) du ,
	\label{eq23}
\end{equation}
and where we are using dimensionless units, following the prescription given in Sec.~\ref{sec22}.
The subsequent evolution of the wave function (\ref{eq22}), which is analytically, is described by
\begin{equation}
	\psi(x,z) = e^{i(x - z^2/6) z/2} Ai(x - z^2/4) .
	\label{eq24}
\end{equation}
As it can be seen,  leaving aside the phase prefactor, the amplitude of this wave function is invariant with the propagation coordinate, $z$, as it is given by an Airy function that moves forward along the transverse coordinate, $x$, as the beam moves ahead in $z$.
This transverse displacement depends quadratically on $z$, which would corresponds to a uniformly accelerated motion in the case of a nonzero mass quantum particle, and that here we thus interpret as a sort of effective transverse acceleration (although we are aware of the fact that the light itself cannot be accelerated).

~

\begin{figure}[t]
	\centering
	\includegraphics[height=11cm]{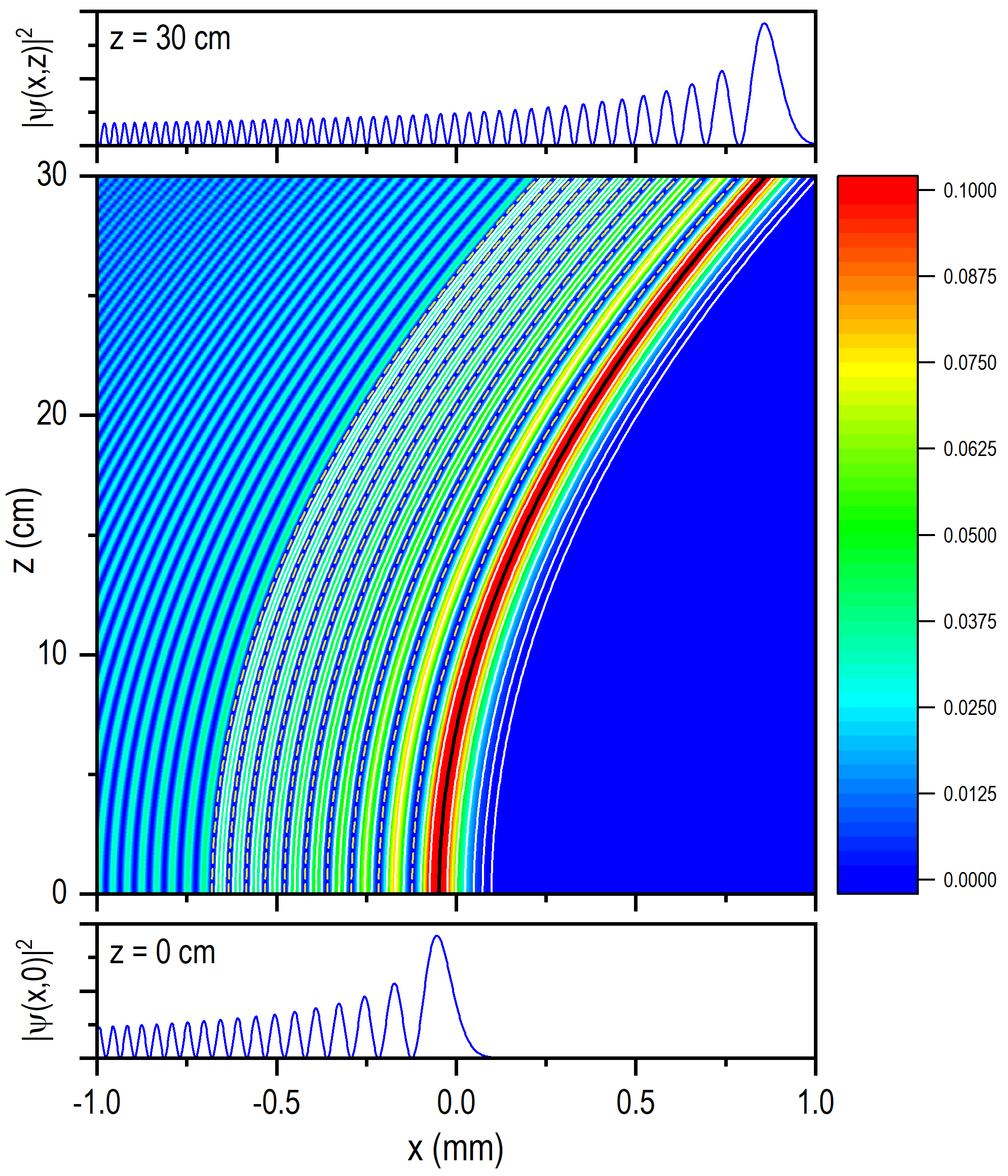}
	\caption{\label{fig1}
		Density plot illustrating the propagation of the intensity distribution associated with an ideal Airy beam.
		A set of Bohmian trajectories (white solid lines) with evenly spaced initial positions covering the first ten maxima is superimposed to the plot in order to stress the preservation of the symmetries of these beams.
		In the graph, both the nodal lines (yellow dashed lines) and the trajectory followed by the center of the main maximum (black solid line) are also shown.
		The profile of the intensity distribution at the input plane ($z=0$~cm) and at $z=30$~cm away are shown below and atop, respectively.
		To compare with experimental data in the finite-energy case, the coordinates are given in the corresponding units (see Ref.~\cite{sanz:JOSAA:2022} for further technical details).}
\end{figure}

Now, the question is how to quantify that, indeed, each part of the wave, and not only the wave as a whole, undergoes a quadratic transverse displacement as the beam propagates forward.
This is where Bohmian trajectories come into play.
From the propagated solution (\ref{eq24}), since the Airy function is real, we obtain that the velocity field only depends on the phase prefactor,
\begin{equation}
	v(x,z) = \frac{dx}{dz} = \frac{z}{2} .
	\label{eq25}
\end{equation}
The integration along $z$ of the effective equation of motion $dx/dz = v(x,z)$ is immediate and renders
\begin{equation}
	x(z) = x(0) + \frac{z^2}{4} .
	\label{eq26}
\end{equation}
We now see that every single piece of the initial intensity distribution propagates following a parabolic trajectory \cite{sanz:JOSAA:2022}, and hence we have an objective element to describe what otherwise is just semi-qualitative.
Note that these trajectories preserve the two transverse symmetries of Airy beams, because they all are parallel one another.
The propagation of an ideal Airy beam is shown in Fig.~\ref{fig1}.
In this figure, a density plot in color scale represents the propagation of the intensity distribution and, superimposed, there is a set of Bohmian trajectories (white solid lines), with evenly spaced initial conditions that cover ten rearmost maxima of the beam.
The intensity profiles at the input ($z=0$) and output ($z=30$~cm) planes are displayed below and atop, respectively.
Also note that both the nodal lines between any two adjacent maxima and the center of the leading maximum are also parallel to the Bohmian trajectories.

~

One may wonder what happens with the above-mentioned symmetries if the beam is close in shape to an Airy function, but it is not an Airy function strictly speaking.
This can happen, for instance, when our imaging system has a limited extension and therefore it is not possible to recreate with a high accuracy, on the input plane, the fast oscillations of the tail of the Airy beam as $x$ gets more and more negative.
Let us thus consider that the beam tail is affected by an exponential decrease \cite{christodoulides:PRL:2007,christodoulides:OptLett:2007}, i.e., the initial field amplitude reads as
\begin{equation}
	\psi(x,0) = Ai(x) e^{\gamma x} .
	\label{eq27}
\end{equation}
This causes that the phase of the Fourier components of the beam, instead of depending on the third power of the transverse momentum,
\begin{equation}
	\tilde{\psi}(k_x) = e^{ik_x^3/3} ,
	\label{eq28}
\end{equation}
as it is the case of the ideal Airy beam, now they are going to show a more complicated dependence,
\begin{equation}
	\tilde{\psi}(k_x) = e^{-\gamma k_x^2 + i(k_x^3 - 3\gamma^2 k_x)/3 + \gamma^3/3} .
	\label{eq29}
\end{equation}
Yet, an analytical expression can still be obtained for the corresponding beam, which reads as
\begin{eqnarray}
	\psi(x,z) & = & e^{i(x - z^2/6) z/2 + \gamma(x - z^2/2) + i\gamma^2 z/2} Ai(x - z^2/4 + i\gamma z) \nonumber \\
	& = & e^{i(x - z^2/6) z/2 + \gamma(y - i\gamma z/2)} Ai(y) ,
	\label{eq30}
\end{eqnarray}
where $y = x - z^2/4 + i\gamma z$.
By virtue of the complex argument $y$, now the Airy function also becomes complex valued.
Therefore, the velocity field will consist of the linear $z$-dependent contribution plus an additional term arising from the phase developed by the Airy function with increasing $z$.
More specifically, this effective velocity field acquires the non-analytical functional form
\begin{equation}
	v(x,z) = \frac{z}{2} + \frac{\partial}{\partial x} \left\{ {\rm Arg} \left[ Ai(y) \right] \right\} ,
\end{equation}
so that the corresponding Bohmian trajectories can only be obtained after numerical integration of the effective equation of motion $dx/dz = v(x,z)$.

~

\begin{figure}[t]
	\centering
	\includegraphics[height=11cm]{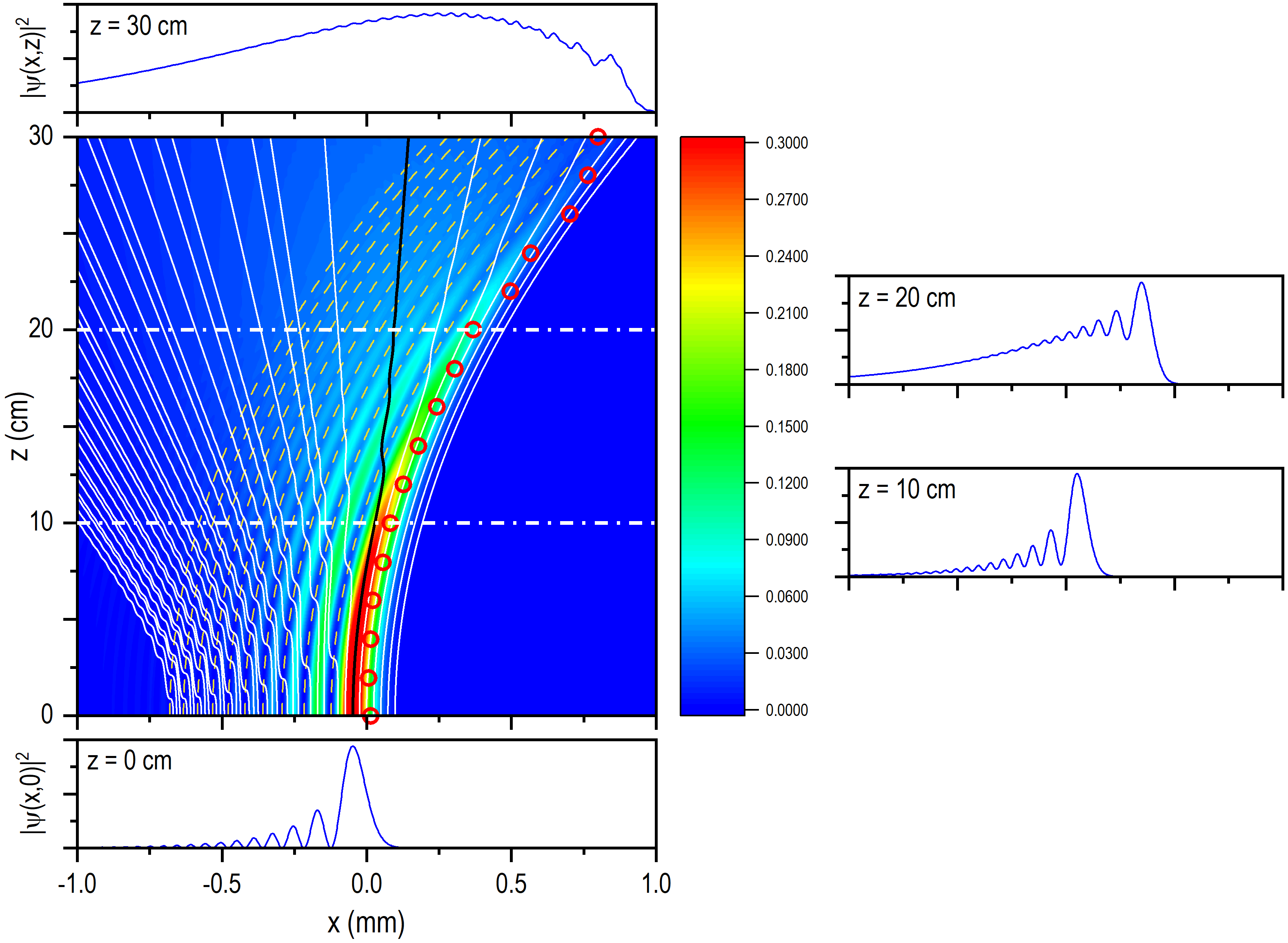}
	\caption{\label{fig2}
		Same as in Fig.~\ref{fig1}, but for a finite energy (non-ideal) Airy beam, with $\gamma = 0.11$.
		To appreciate how the Bohmian trajectories (energy flux) deviates from its ideal behavior and hence the typical Airy pattern gradually disappears, giving rise to a Gaussian type profile (see atop intensity profile), the nodal lines of the ideal case are also included (yellow dashed lines).
		On the side panels, density profiles at $z = 10$~cm and $z=20$~cm away from the input plane.
		The red circles correspond to the experimental data reported in \cite{christodoulides:PRL:2007} (see Ref.~\cite{sanz:JOSAA:2022} for further technical details).}
\end{figure}

Figure~\ref{fig2} shows the propagation of an Airy bean with exponentially decreasing tail, also as in Fig.~\ref{fig1}, in terms of a color-scaled density plot of the intensity distribution, where a series of intensity profiles are shown on the side panels for $z = 0$, 10~cm, 20~cm, and 30~cm (see horizontal dashed lines in density plot).
Moreover, the Bohmian trajectories corresponding to exactly the same initial conditions as in Fig.~\ref{fig1} are also shown superimposed.
Note that, while the density plot gradually loses the features that characterized the ideal behavior, the trajectories provide us with precise information on how such a loss process is taking place, namely, by letting the intensity to be gradually transferred from one maximum to the immediately adjacent one from behind.
This is possible, because there is no more light intensity coming from the rearmost part of the beam, pressing ahead, and hence, as it happens with a water stream, the intensity starts flowing backwards.
The trajectories only indicate how the backwards flow is taking place at each point as the beam moves forward.
On top of the plot, we have also included the experimental data reported by Siviloglou {\it et al.}~\cite{christodoulides:PRL:2007}, showing the excellent agreement with ours.

%%%%%%%%%%%%%%%%%%%%%%%%%%%%%%%%%%%%%%%%%%%%%%%%%%%%%%%%%%%%%%%%%%%%%%%%
%%%%%%%%%%%%%%%%%%%%%%%%%%%%%%%%%%%%%%%%%%%%%%%%%%%%%%%%%%%%%%%%%%%%%%%%

\section{Spontaneous self-focusing}
\label{sec4}

In the same way that converging lenses produce light focusing at a given distance behind them when plane waves are incident on them \cite{aiello:OptLett:2016,aiello:OSACont:2020,porras:PRA:2021}, a similar behavior can also be found with quantum systems.
In his renowned monograph {\it Quantum Theory: Concepts and Methods}, Asher Peres put forth \cite{peres-bk} the possibility to have, at the same time, wave functions that are consistent with Schr\"odinger's equation and singular at a given time.
Specifically, he conjectured that:
\begin{quotation}
	It is inconsistent to require Schr\"odinger wave functions to be always  continuous and finite, even for free particles. It is not difficult to construct states which are represented, at time $t = 0$, by a continuous function and which evolve into a discontinuous, or even singular, one.
\end{quotation}
To illustrate this fact, Peres considers an initial ansatz of the type
\begin{equation}
	\psi(x,0) = \frac{e^{-i x^2/2}}{(1 + x^2)^{1/3}} ,
	\label{eq32}
\end{equation}
where we use $x$ and $z$ in order to continue with the optical analogy (in the quantum case, $\hbar = m = 1$; in the optical case, $k=1$).
The free-space propagation of the ansatz (\ref{eq32}) leads to the field amplitude at a distance $z$
\begin{equation}
	\psi(x,z) = \frac{e^{i x^2/2z}}{\sqrt{2\pi iz}} \int_{-\infty}^\infty \frac{e^{i(z_f-z) x'^2/2z - ixx'/z}}{(1+x'^2)^{1/3}} dx' ,
	\label{eq33}
\end{equation}
with $z_f = 1$ being the focusing distance.
At this latter distance, as it was pointed out by Peres, this solution presents a singularity at the origin ($x=0$), even though it satisfies Schr\"odinger's equation, due to the divergence at that point of the integral in
\begin{equation}
	\psi(x,z_f) = \frac{e^{ix^2/2}}{\sqrt{2\pi iz_f}}\
	\int_{-\infty}^\infty \frac{e^{-ixx'/z_f}}{(1 + x'^2)^{1/3}} dx' .
	\label{eq34}
\end{equation}
This solution actually belongs to a wider class of solutions, which are described in detailed, as well as the associated blowing-up effect at the origin, in Ref.~\cite{sanz:PLA:2024}.

\begin{figure}[!t]
	\centering
	\includegraphics[width=15cm]{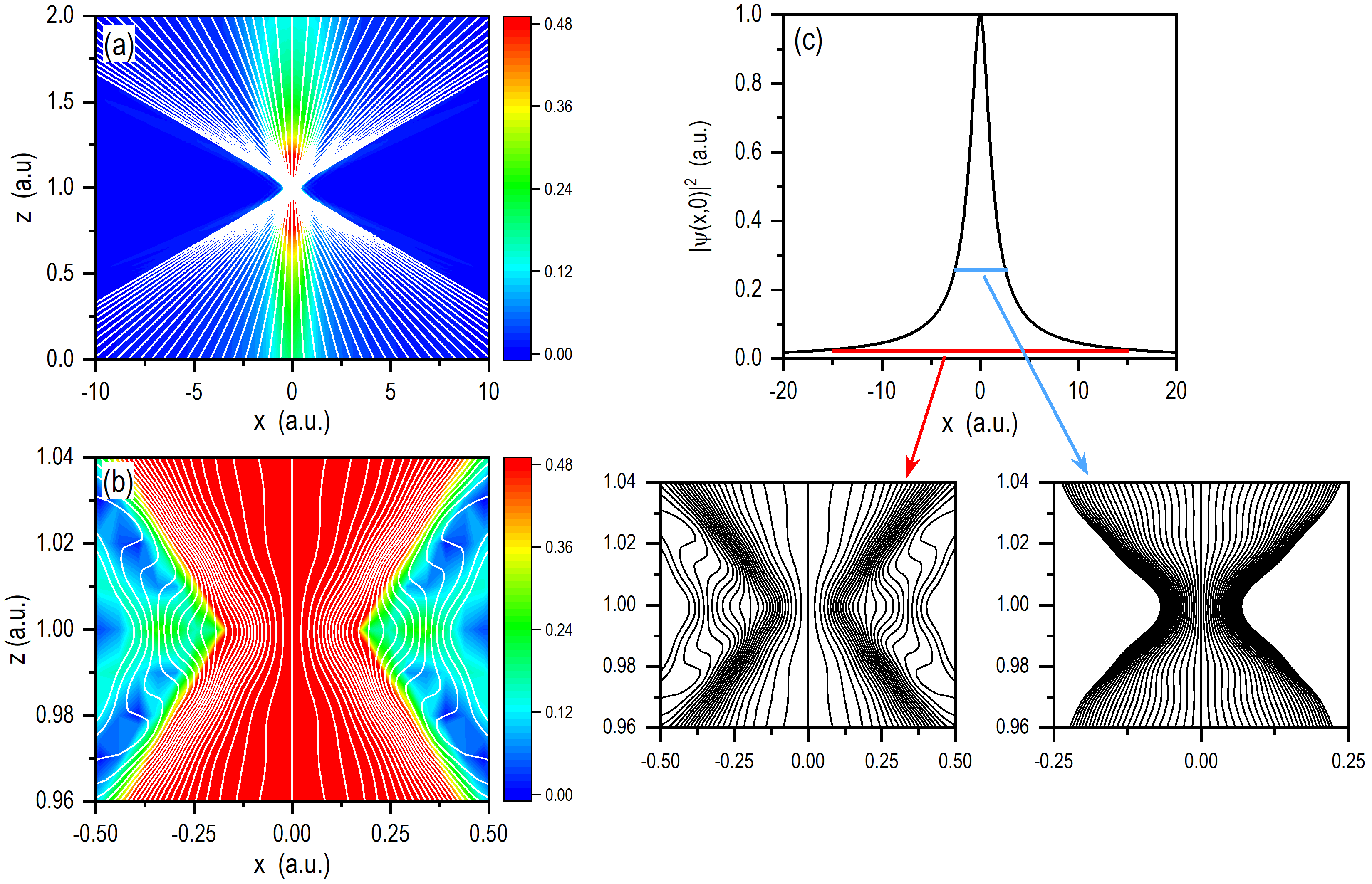}
	\caption{\label{fig3}
		(a) Density plot showing the propagation and self-focusing effect of the Peres ansatz, Eq.~(\ref{eq32}).
		A set of 51 Bohmian trajectories with evenly spaced initial positions appears superimposed.
		To better appreciate the self-focusing effect around $z_f$ in (b), panel (b) shows an enlargement of the focal region.
		(c) Intensity distribution for the Peres ansatz.
		Each horizontal line denotes a region over which a set of evenly spaced initial positions has been considered for the sets of trajectories shown below, where it is seen that, as they are picked up closer to the central part of the beam, the focusing is more apparent.
		In all plots, arbitrary units (a.u.) are considered.}
\end{figure}

~

A numerical simulation of this phenomenon is shown in Fig.~\ref{fig3}(a), where we can observe how a set of 51 Bohmian trajectories superimposed to the density plot of the beam intensity quickly bent over the focal region, around $x=0$, at $z_f$.
In order to better appreciate the behavior of the trajectories around the focal region, an enlargement is shown in Fig.~\ref{fig3}(b), where we can also observe the distortion of the seemingly straight rays in panel (a) due to the presence of nodes of the beam amplitude.
The initial positions of the trajectories have been chosen evenly distributed in $x$ covering the range $(-15,15)$, which basically spans the basis of the intensity distribution (truncated at the borders to avoid numerical instabilities \cite{sanz:PLA:2024}), as seen in Fig.~\ref{fig3}(c), where such a range is denoted with a red line, at about $2\%$ the maximum.
This might generates a misleading impression that, in fact, the trajectories try to avoid the focus $x=0$.
However, if the initial positions are taken closer to the focus, at about $30\%$ the maximum intensity [blue line in Fig.~\ref{fig3}(c)], we observe in Fig.~\ref{fig3}(d) how the corresponding trajectories actually accumulate around the focus.

~

Taking into account the similarity between the ansatz (\ref{eq32}) and a Gaussian beam around the center of the intensity distribution displayed in Fig.~\ref{fig3}(c), we can use the latter to understand this type of inversion symmetry with respect of the focus, which resembles the effect of a converging lens in geometrical optics.
To this end, consider an initial Gaussian beam written in the form
\begin{equation}
	\psi(x,0) = e^{-x^2/4\sigma_0^2} ,
	\label{eq35}
\end{equation}
with a free propagation that can be recast as
\begin{equation}
	\psi(x,z)
	= \sqrt{\frac{\sigma_0}{\tilde{\sigma}_{z}}} e^{-x^2/4\sigma_0\tilde{\sigma}_{z}}
	= \sqrt{\frac{\sigma_0}{\tilde{\sigma}_{z}}} e^{-x^2/4\sigma_{z}^2} e^{i x^2 z/8\sigma_0^2 \sigma_{z}^2} .
	\label{eq36}
\end{equation}
In the first term, the explicit expression for the complex width parameter is
\begin{equation}
	\tilde{\sigma}_{z} = \sigma_0 \left( 1 + \frac{iz}{2\sigma_0^2} \right)
	\label{eq37}
\end{equation}
which accounts for both the dispersion of the wave packet during its propagation along $z$, given by
\begin{equation}
	\sigma_{z} = |\tilde{\sigma}_{z}|
	= \sigma_0 \sqrt{1 + \left( \frac{z}{2\sigma_0^2} \right)^2} ,
	\label{eq38}
\end{equation}
and the development of a space-dependent phase factor, as it is explicitly seen behind the second equality in (\ref{eq36}).
The associated Bohmian trajectories can readily be obtained analytically \cite{sanz-bk-2} from the equation of motion
\begin{equation}
	\dot{x} = \left( \frac{z}{2\sigma_0^2} \right)^2
	\frac{\sigma_0^2}{\sigma_z^2} x ,
	\label{eq39}
\end{equation}
which renders
\begin{equation}
	x(z) = \frac{\sigma_{z}}{\sigma_0} x(0) .
	\label{eq40}
\end{equation}
From the latter expression, we note that the focus or waist of the beam takes place at $z=0$, from which all trajectories diverge symmetrically forward (for $z>0$) or backwards ($z<0$) as hyperbolas.
Actually, for rather large values of $z$, the trajectories behave as straight rays, just as in the case of the Peres' ansatz.

~

In order to find a behavior closer even to the Peres' ansatz, which can help us to understand the focusing effect, we need to make the Gaussian beam to focus at $z_f$ instead of $z=0$.
This can be achieved by considering the generalized complex width parameter
\begin{equation}
	\tilde{\sigma}_{g,z}
	= \sigma_0 \left[ 1 + \frac{i (z - z_f)}{2\sigma_0^2} \right]
	= \tilde{\sigma}_{g,0}
	\left( 1 + \frac{i z}{2\sigma_0\tilde{\sigma}_{g,0}} \right) ,
	\label{eq41}
\end{equation}
with
\begin{equation}
	\tilde{\sigma}_{g,0} =
	\sigma_0 \left( 1 - \frac{i z_f}{2\sigma_0^2} \right) .
	\label{eq42}
\end{equation}
Accordingly, the width $\sigma_0$ of the initial Gaussian ansatz (\ref{eq35}) has to be replaced by $\tilde{\sigma}_{g,0}$, which physically means that the observation of spontaneous focusing is directly related to imprinting a phase in the initial beam.
That is, we explicitly observe that the phase plays a major role in these dynamics, more prominent, actually, than the shape (amplitude) itself of the beam.
Taking this into account, note that the width and phase of the initial value $\tilde{\sigma}_{g,0}$ are, respectively,
\begin{eqnarray}
	\sigma_{g,0} & = &
	\sigma_0 \sqrt{ 1 + \left( \frac{z_f}{2\sigma_0^2} \right)^2} ,
	\label{eq43} \\
	\theta_{g,0} & = & \left( \tan \right)^{-1}   \left( - \frac{z_f}{2\sigma_0^2} \right) ,
	\label{eq44}
\end{eqnarray}
which, at any other value of $z$, read as
\begin{eqnarray}
	\sigma_{g,z} & = & \sigma_0 \sqrt{1 + \left( \frac{z - z_f}{2\sigma_0^2} \right)^2} ,
	\label{eq45} \\
	\theta_{g,z} & = & \left( \tan \right)^{-1} \left( \frac{z - z_f}{2\sigma_0^2} \right) .
	\label{eq46}
\end{eqnarray}
This latter expressions show, precisely, that at $z=z_f$, the Gaussian beam corresponds to the ansanzt (\ref{eq36}), with minimum width $\sigma_{g,z_f}=\sigma_0$ and zero phase, $\theta_{g,z_f}=0$.

\begin{figure}[!t]
	\centering
	\includegraphics[width=12cm]{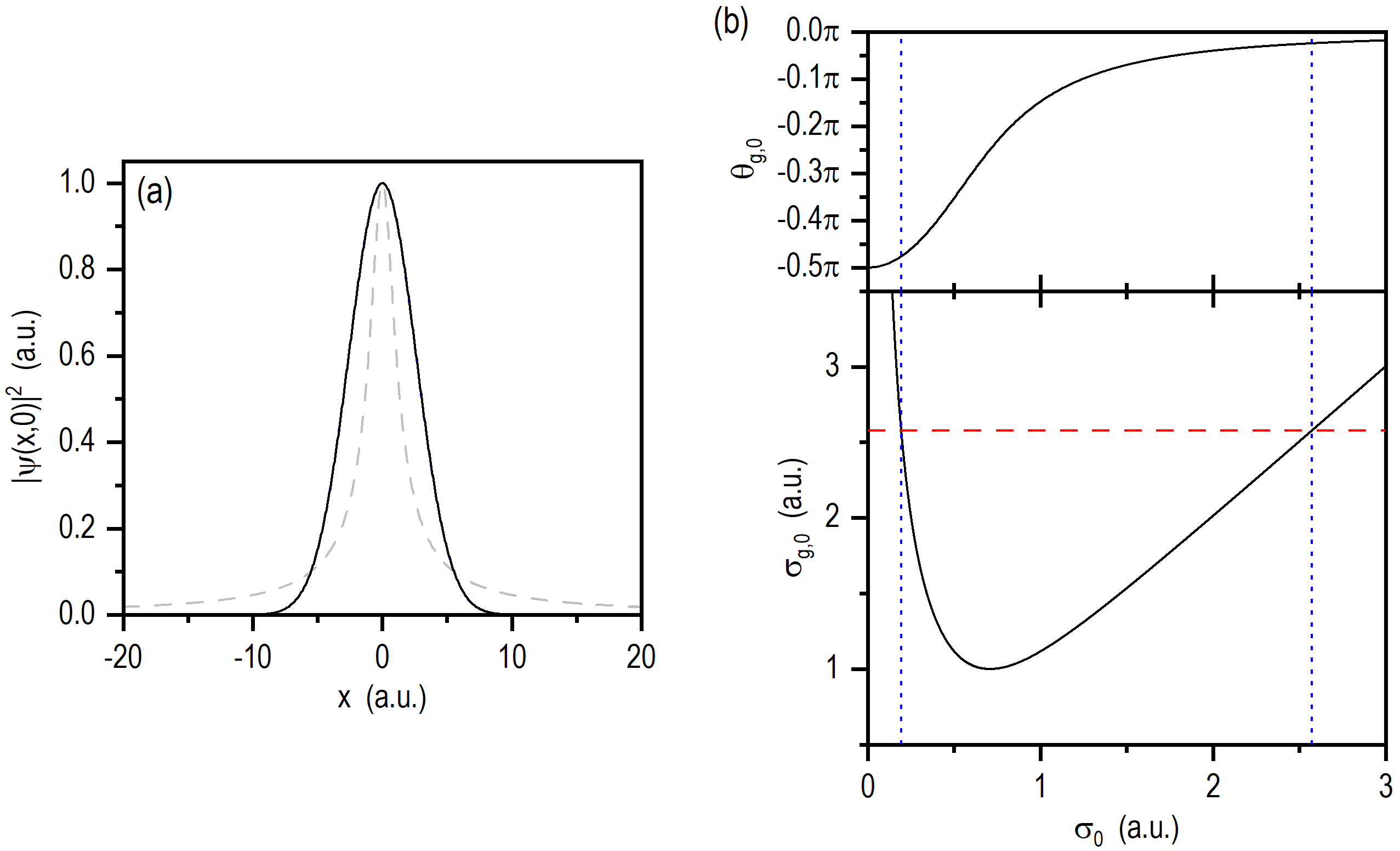}
	\caption{\label{fig4}
		(a) Intensity distribution at $z = 0$ for a Gaussian beam with its width at $10\%$ its maximum value being equal to the corresponding value of the intensity distribution corresponding to the Peres ansatz, Eq.~(\ref{eq32}), which, for comparison purposes is represented with gray dashed line.
		(b) Initial width (bottom), Eq.~(\ref{eq43}), and phase (top), Eq.~(\ref{eq44}), of the generalized complex width parameter, $\tilde{\sigma}_{g,0}$, as a function of the waist width $\sigma_0$.
		The red dashed line in the plot for $\sigma_{g,0}$ illustrates that, for the same value of this quantity, there are two possible values of $\sigma_0$.
		In this particular case, such values are $\sigma_{0,+} \approx 2.571$ and $\sigma_{0,-} \approx 0.194$, which are identified by the vertical blue dotted lines.
		These two values correspond, respectively,to the initial phases $\theta_{g,0}^+ \approx -0.024\pi$ versus $\theta_{g,0}^- \approx -0.477\pi$.
		In all plots, arbitrary units (a.u.) are considered.}
\end{figure}

~

\begin{figure}[!t]
	\centering
	\includegraphics[width=12cm]{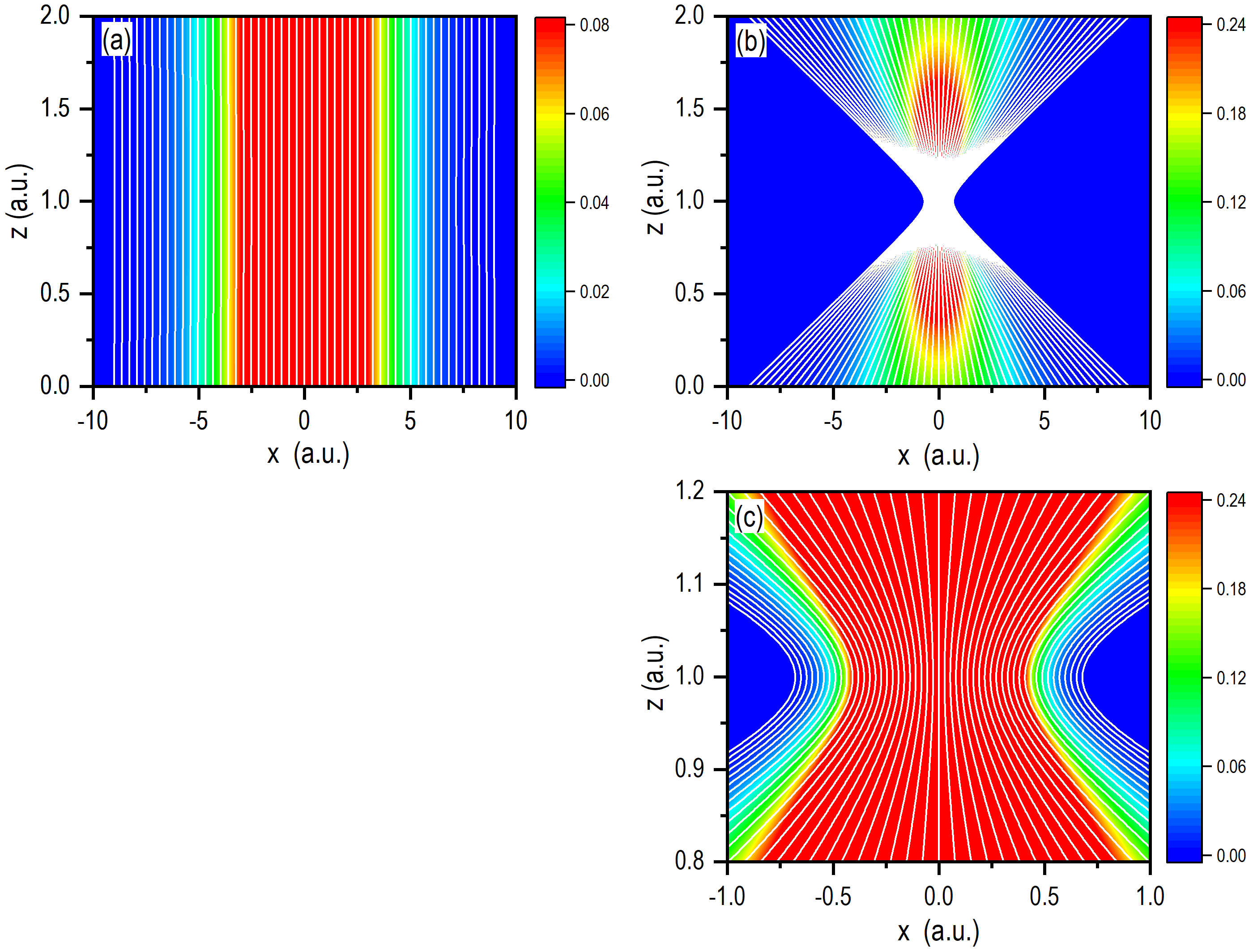}
	\caption{\label{fig5}
		Density plots showing the propagation of Gaussian beams with the same initial width, but different phase: (a) $\sigma_{0,+} \approx 2.571$ and (b/c) $\sigma_{0,-} \approx 0.194$.
		In each plot, a set of 51 Bohmian trajectories with evenly spaced initial positions has been superimposed (the initial positions are the same in the two cases).
		To better appreciate the self-focusing effect around $z_f$ in (b), an enlargement of the focal region is shown in panel (c).
		In all plots, arbitrary units (a.u.) are considered.}
\end{figure}

Now, unlike the standard case, where $z_f=0$, by virtue of Eq.~(\ref{eq43}) the same initial value $\sigma_{g,0}$ may arise from two different values of $\sigma_0$:
\begin{equation}
	\sigma_{0,\pm} = \sqrt{\frac{\sigma_{g,0}^2 \pm \sqrt{\sigma_{g,0}^4 - z_f^2}}{2}} ,
	\label{eq47}
\end{equation}
for $\sigma_{g,0}^2 > z_f$ (for $\sigma_{g,0}^2 = z_f$, the solution is degenerate, i.e., $\sigma_{0,+} = \sigma_{0,-}$, while for lower values of $z_f$ there are no solutions).
To illustrate this situation, let us consider an initial Gaussian beam with its value at a tenth its maximum [that is, $|\psi(x_\pm,0)|^2/|\psi(0,0)|^2 = 0.1$] being equal to the corresponding value of the Peres field amplitude (\ref{eq32}), as seen in Fig.~\ref{fig4}(a).
This renders
\begin{equation}
	\sigma_{g,0} = \sqrt{\frac{10\sqrt{10}-1}{2 \ln 10}} \approx 2.579 ,
	\label{eq48}
\end{equation}
from which, after substituting it into Eq.~(\ref{eq47}), we obtain $\sigma_{0,+} \approx 2.571$ and $\sigma_{0,-} \approx 0.194$, as seen in Fig.~\ref{fig4}(b).
Note that, while in the first case we do not expect a remarkable dynamical behavior, since $\sigma_{0,+} \sim 99\% \sigma_{g,0}$ at $z = z_f$, in the second case, $\sigma_{0,-} \sim 7.5\% \sigma_{g,0}$, which means that there is a rather important spontaneous self-focusing effect.
As it is shown in Fig.~\ref{fig4}(a), although at $z=0$ both Gaussian beams look the same regarding their intensity distribution, both with the same width, as it is denoted by the horizontal red dashed line of the lower panel of Fig.~\ref{fig4}(b), the initial phase of their complex width parameters is quite different.
In the case of $\sigma_{0,+}$, the associated phase is $\theta_{g,0}^+ \approx -0.024\pi$, which is negligible and, therefore, no relevant dynamical effects should be expected.
On the other hand, for $\sigma_{0,-}$, we obtain $\theta_{g,0}^- \approx -0.477\pi$, close to $-\pi/2$, and hence we might expect an important effect in the dynamics.
This is, precisely, what we observe in Figs.~\ref{fig5}(a) and \ref{fig5}(b), for $\sigma_{0,+}$ and $\sigma_{0,-}$, respectively: while in the first case, the trajectories appear as nearly straight rays with almost no deflection, in the second case there is an important focusing effect at $z=z_f$ [an enlargement around the focusing region in the latter case is shown in Fig.~\ref{fig5}(c) to better appreciate the strong compression of the ray beam at the focal region].
The trajectories in either case have been obtained analytically, from the trajectory equation
\begin{equation}
	x(z) = \frac{\sigma_{g,z}}{\sigma_{g,0}} x(0) ,
	\label{eq49}
\end{equation}
which arises after following the same procedure as in the conventional case, but replacing $\tilde{\sigma}_{z}$ by $\sigma_{g,z}$.
Note that these trajectories are also hyperbolas with their minimum deflection at the beam waist, which, in this case, coincides with the focal region.

%%%%%%%%%%%%%%%%%%%%%%%%%%%%%%%%%%%%%%%%%%%%%%%%%%%%%%%%%%%%%%%%%%%%%%%%
%%%%%%%%%%%%%%%%%%%%%%%%%%%%%%%%%%%%%%%%%%%%%%%%%%%%%%%%%%%%%%%%%%%%%%%%

\section{Concluding remarks}
\label{sec5}

The main purpose of this Communication was to emphasize the role of the so-called Bohmian velocity field to interpret and understand the spatial dispersion of waves, either quantum mechanical wave packets or optical light beams, which gives rise to the corresponding (spatial) distributions of the probability density, in the first case, or the light intensity (electromagnetic energy density), in the second one.
We have seen that, beyond ontological issues, this field is well-defined from a formal point of view within conventional quantum mechanics.
Indeed, it is a direct consequence, from a physical point of view, of the local (spatial) phase variations, which thus allows us to obtain and evaluate global information about the diffusion dynamics in the configuration space.
The same field, but introduced in an effective manner, can also be appealed to in other theories described by Schr\"odinger-type equations, as it is the case of wave optics, whenever the light fields are describable by means of the paraxial Helmholtz equation, although a more rigorous treatment, based on Maxwell's equations and the use of the Poynting vector, also renders, when the paraxial approximation is invoked, the same result \cite{sanz:ApplSci:2020}.
This is just a reminiscence of Hamilton's optical-mechanical analogy \cite{bornwolf-bk}, later on used by Schr\"odinger to derive his formulation of wave mechanics \cite{schrodinger:PhysRev:1926}.
Of course, in the optical case here investigated, it is important to note that the propagation variable is not time, but the longitudinal coordinate (the $z$-coordinate, here), and that the transverse plane is the playground where the important physics takes places, instead of the configuration space.

~

We have also seen that, in order to determine the flow of probability or light intensity in space, a rather convenient tool is provided by the trajectories that arise after integrating along the propagation variable the above-mentioned velocity field.
This renders the so-called Bohmian trajectories, which, again, have nothing to do (at least, in principle) with ontological hidden variables, but just with well-defined elements naturally arising within a hydrodynamic reformulation of Schr\"odinger and, in general, Schr\"odinger-type equations.
These trajectories allow us to extract valuable local information about the propagation of the systems described by such equations, which, in turn, supply an additional perspective on how and why probabilities or light intensities redistribute spatially in the way they do it.
This event-by-event picture of wave phenomena results worth emphasizing, because it enables the reconstruction of such phenomena under a single-event basis (single photons, single electrons, etc.), that is, in circumstances of a rather low flux of particles \cite{weis:AJP:2008,padgett:AJP:2016,pozzi:EJP:2013,batelaan:NJP:2013}, thus making clearer the connection between particle wave theories and statistics \cite{sanz:AnnFondLdB:2021}.

~

Here, in particular, we have considered these tools to study and explain the symmetries and singularities that characterize two types of challenging behaviors that arise in the simplest propagation conditions, i.e., under free propagation.
Specifically, we have investigated those properties in two phenomena that have been recently considered in more detail elsewhere: ($i$) the lack of dispersion and self-acceleration of Airy beams during their propagation \cite{sanz:JOSAA:2022}, and ($ii$) the spontaneous self-focusing displayed by a beam when a particular phase distribution is imprinted in it initially \cite{sanz:PLA:2024}.
In the first case, the trajectories have allowed us to provide a quantitative measure to understand how Airy-type beams degrade when we introduce conditions that diminish their ideal initial shape.
Thus, we find that, the translation symmetry along the hyperbolic Bohmian trajectories disappears gradually, because there is an effective transfer of flux towards the rear part of the beam, which eventually translates also into a loss of its transverse self-acceleration.
In other words, what keeps the Airy beam properties is the presence of an infinite tail.
This generates a constant pressure forwards \cite{takabayasi:ProgTheorPhys:1952,takabayasi:ProgTheorPhys:1953}, which is evidenced by all trajectories, regardless of their initial position, bending over at a constant rate.

~

In the case of the spontaneous self-focusing (i.e., without the mediation of nonlinear terms in the Schr\"odinger equation), we have observed that it is characterized, in a good approximation, by both mirror symmetry  with respect to the propagation axis and time-reversal symmetry with respect to the focal region, where the singularity appears.
This behavior is analogous to the focusing produced by a converging lens when a plane wave is incident on it, where the geometrical ray description coincides pretty well with the trajectories here obtained.
Nonetheless, unlike geometrical rays, Bohmian trajectories cannot cross through the same point for the same value of the propagation variable, which leads to a bouncing backward motion after they reach the focal region.
More importantly, we have observed that the focusing affecting the ensemble of trajectories (and hence of the beam) is directly related to the addition or imprinting of an extra phase, which is in compliance with the fact that it is a velocity field (generated by the local phase variations of the wave) what causes that probability densities or light intensities behave in the way they do.

%%%%%%%%%%%%%%%%%%%%%%%%%%%%%%%%%%%%%%%%%%%%%%%%%%%%%%%%%%%%%%%%%%%%%%%
%%%%%%%%%%%%%%%%%%%%%%%%%%%%%%%%%%%%%%%%%%%%%%%%%%%%%%%%%%%%%%%%%%%%%%%

\section*{Acknowledgments}

%The author thanks the organizers of the International Symposium {\it Symmetries in Science XIX} (Bregenz, June 30 -- August 4th, 2023), as well as Dieter and Ivette, for their invitation to participate in the event and their kind hospitality while in Bregenz.
%Part of the work presented here arises from a fruitful collaboration with Luis L.~S\'anchez-Soto and Andrea Aiello, whom he is also thankful for the interesting discussions that eventually gave rise to Ref.~\cite{sanz:PLA:2024}.
Financial support from the Spanish Agencia Estatal de Investigaci\'on (AEI) and the European Regional Development Fund (ERDF) (Grant No. PID2021-127781NB-I00) is acknowledged.

\end{document}